\documentclass[]{article}

\usepackage{amsmath,amssymb,amsfonts,amsthm,mathrsfs,paralist}
\usepackage{mathtools,enumitem,color}
\usepackage{setspace}
\usepackage[dvipsnames]{xcolor}
\theoremstyle{plain}
\newtheorem{thm}{Theorem}[section]

\newtheorem{lemma}[thm]{Lemma}
\newtheorem{cor}[thm]{Corollary}
\usepackage{subcaption}

\theoremstyle{definition}
\newtheorem{defi}[thm]{Definition}

\theoremstyle{remark}

\newcommand{\N}{\mathbb {N}}

\newcommand{\dg}[1]{{\color{black}{#1}}}

\def\BFa {\mathbf{a}}

\def\BFb {\mathbf{b}}

\def\BFu {\mathbf{u}}
\def\BFU {\mathbf{U}}

\def\BFv {\mathbf{v}}
\def\BFx {\mathbf{x}}

\def\BFy {\mathbf{y}}

\def\BFi {\textit{\bfseries i}}
\def\BFk {\textit{\bfseries k}}

\usepackage[ruled,vlined]{algorithm2e}

\def\hgracia#1{}
\def\hchristiane#1{}
\def\hhenri#1{}

\def\statement{\showfigures}

\newcommand{\hidefigures}[1]{}
\newcommand{\showfigures}[1]{#1}

\begin{document}


  \title{A negative dependence framework to assess different forms of scrambling }

  \author{Henri Faure\footnote{I2M, Institut de Math\'ematiques de Marseille, France, henri.faure@univ-amu.fr},
    Gracia Y.\ Dong\footnote{Department of Statistics and Actuarial Science, University of
    Waterloo, Canada, gracia.dong@uwaterloo.ca} and Christiane Lemieux\footnote{Department of Statistics and Actuarial Science, University of
    Waterloo, Canada, clemieux@uwaterloo.ca}
  }
  
  \date{\normalsize\today}

  \maketitle

\begin{abstract}
We use the framework of dependence to assess the benefits of scrambling randomly versus deterministically for Faure and Halton sequences. We attempt to answer the following questions: when a deterministic sequence has known defects for small sample sizes, should we address these defects by applying random scrambling or should we find a ``good'' deterministic scrambling yielding a sequence that can then be randomized using a less computer-intensive randomization method such as a digital shift? And in the latter case, how do we choose a deterministic scrambling and how do we assess whether it is good or not?
\end{abstract}

\section{Introduction}

Low-discrepancy sequences such as the Halton, Faure and Sobol' sequences \cite{rHAL60a, faure1982,sobol1967distribution} can provide estimators for high-dimensional integration problems that are more accurate than those obtained using a sequence of random points as in the Monte Carlo method. Their use in this context is often referred to as {\em quasi-Monte Carlo methods}.

The Faure and Halton sequences along with some of the generalizations that have been proposed over the years \cite{rTEZ94b,rTEZ95a,faure2009generalized} stand out among low-discrepancy sequences thanks to the fact that they achieve a certain form of perfect equidistribution. In the case of the Faure sequence, which is a special case of a digital $(t,s)-$sequence in base $b$, it is captured by the quality parameter $t$ which can be shown to be 0 in this case \cite{faure1982}. The Halton sequence is not a digital $(t,s)$-sequence as it is constructed using a different paradigm requiring for a different base $b$ for each coordinate of the points. However, a similar notion of perfect equidistribution holds for this construction, in the sense that for specific partitions of the unit hypercube $[0,1)^s$ in which they lie, one can prove that certain segments of the sequence will put the same number of points in each subset from the partition. 

While these sequences achieve an optimal equidistribution as described above, this optimal behaviour can require a very large number of points before being observed, depending on the bases used, which in turn depends on the dimension $s$ of the sequence. For larger bases, if we use the first $n$ points of the sequence and $n$ is too small compared to the number of points where the optimal equidistribution properties are shown to hold, the corresponding point set may not be of very good quality, thus yielding quasi-Monte Carlo estimators that are potentially less accurate than those obtained using the Monte Carlo method.

To alleviate this problem, several authors have proposed to apply certain types of {\em scrambling} to the Faure and Halton sequences. Some early work in this area goes back to Faure \cite{faure1992good}, who proposed to use permutations of the integers in $[0,\ldots,b-1]$ for the van der Corput sequence in base $b$. Following this work, many other deterministic scramblings based on permutations have been proposed for the Halton sequences, see \cite{faure2009generalized, donglemieux22} and the references therein. 
Similarly, for the Faure sequences, generalizations of the original construction from \cite{faure1982} have been proposed, starting with Tezuka \cite{rTEZ94b} and Tezuka-Tokuyama \cite{qFAU09a}. Further \cite{rTEZ94b} was developed by Papageorgiou and Traub in a software system called Finder, see 
\cite{BeatMC}, widely used in finance. More recently, new perspectives on 
Faure sequences have been explored in \cite{qFAU09a}, getting new generating matrices.

In addition, random scramblings have been proposed for both digital $(t,s)$-sequences \cite{owen1995randomly,matousek1998thel2} and the Halton sequence \cite{donglemieux22,okten2009generalized,owen2017randomized}. Both deterministic and random scramblings are widely accepted as providing  an improvement to the original constructions from \cite{faure1982,rHAL60a} especially for small number of points in medium to large dimensions.


An important question faced by anyone who wants to use these constructions is whether a deterministic or a random scrambling should be applied. On one hand, one would think that a very well chosen deterministic scrambling should do better than one randomly chosen, and would have a less computationally demanding implementation since it could then be used with a ``cheap'' randomization such as a digital shift for the purpose of error estimation. On the other hand, finding a good deterministic scrambling that performs well on a large variety of problems may not be possible, and from that point of view, perhaps random scramblings are a better option.

In this paper, we aim to answer this question using the framework of negative dependence introduced in \cite{donglemieux22,Lem17,wiart2021dependence}. On one hand, results in these papers have demonstrated that the optimal equidistribution of the Faure and Halton sequence is a sufficient and necessary condition for random scrambling to induce a certain form of negative dependence. This means that for these sequences, random scrambling can ``repair'' whatever defects these sequences may have in their initial portion. This framework also yields a quality measure that can capture some of the equidistribution properties of point sets that are missed by the quality parameter $t$. Hence we propose to use this framework to assess deterministic scramblings, which has the advantage of providing a common framework to study and compare random and deterministic scramblings, for both Halton and Faure sequences. We also examine how the permutations from \cite{faure1992good} can be used not only for Halton sequences but also for Faure sequences, thereby proposing a new form of deterministic scramblings for the latter. Numerical experiments comparing deterministic and random scramblings are performed on a variety of problems to provide empirical evidence toward answering our main question.

\section{Negative dependence framework}
\label{sec:negdep}


In this section we introduce definitions and results from \cite{Lem17,wiart2021dependence,donglemieux22} that are needed to describe the negative dependence framework used later on to assess different scrambling approaches.

It should not come as a surprise to the reader that in order to use dependence concepts to assess point sets, we need to study {\em pairs} of points and study their behavior compared to pairs of independently sampled points. The following definition introduces a key quantity needed to study the behavior of pairs of points.

\begin{defi}
        For $x,y\in[0,1)$, let $\gamma_b(x,y) \ge 0$ be the exact number of initial digits shared by $x$ and $y$ in their base $b$ expansion, i.e.\ the smallest number $i \ge 0$ such that
	\begin{equation} \label{eq:gammadef}
	\lfloor b^ix\rfloor=\lfloor b^iy\rfloor\quad\text{but}\quad\lfloor b^{i+1}x\rfloor \neq\lfloor b^{i+1}y\rfloor.
	\end{equation}
	\hchristiane{added following sentence and put two $\ge 0$ above. I don't think we need to explain that yes indeed we can have $\gamma_b(x,y)=0$.}
	If $x=y$ then we let $\gamma_b(x,y) = \infty$.
%
	For $\BFx,\BFy\in[0,1)^s$, we define 
	\[
	\boldsymbol{\gamma}^s_b(\BFx,\BFy) = (\gamma_b(x_1,y_1),\ldots,\gamma_b(x_s,y_s))
	\mbox{ and }
	\gamma_b(\BFx,\BFy) = \sum_{j=1}^s \gamma_b(x_j,y_j).
	\]	
	For $\BFb=(b_1,\ldots,b_s)$ we define
		\[
	\boldsymbol{\gamma}^s_{\BFb}(\BFx,\BFy) = (\gamma_{b_1}(x_1,y_1),\ldots,\gamma_{b_s}(x_s,y_s))
	\mbox{ and }
	\gamma_{\BFb}(\BFx,\BFy) = \sum_{j=1}^s \gamma_{b_j}(x_j,y_j).
	\]
	\hchristiane{added following sentence}
\end{defi}

Note that $\boldsymbol{\gamma}^s_b(\BFx,\BFy)$ and $\boldsymbol{\gamma}^s_{\BFb}(\BFx,\BFy)$ denote  $s$-dimensional vectors while 	$ \gamma_b(\BFx,\BFy)$ and $ \gamma_{\BFb}(\BFx,\BFy)$ are scalars.  Also, note that \eqref{eq:gammadef} implies  $\gamma_b(x,y)$ is well defined for any $x,y \in [0,1)$ even if  $x,y$ do not have a unique expansion in base $b$.

Given $\BFi,\BFk \in \mathbb{N}^s$, we say that $\BFk \le \BFi$ if $k_j \le i_j$ for all $j=1,\ldots,s$. (Note that in this paper, we assume that $\N$ includes 0.) 
We also denote the $\ell_1$-norm of a vector $\BFk$ by 
$|\BFk|=k_1 + \ldots + k_s$. 

Next we introduce counting numbers from \cite{donglemieux22, wiart2021dependence} that are needed to define the quality measure $C_b(\BFk;P_n)$ that will be crucial in our analysis.

\begin{defi}\label{def:mknipn}
Let $P_n =\{\BFU_1,\ldots,\BFU_n\}$ be a point set in $[0,1)^s$ and $b \ge 2$ be an integer. 
%
Let $M_b(\BFk;P_n)$ be the number of ordered pairs of distinct points $(\BFU_l,\BFU_j)$ in $P_n$ such that $\boldsymbol{\gamma}_b^s(\BFU_l,\BFU_j) \ge \BFk$. Similarly, we let $M_{\BFb}(\BFk;P_n)$ be the number of ordered pairs of distinct points $(\BFU_l,\BFU_j)$ in $P_n$ such that $\boldsymbol{\gamma}_{\BFb}^s(\BFU_l,\BFU_j) \ge \BFk$.
\end{defi}

We can connect the quantity $M_b(\BFk;P_n)$ with the concept of equidistribution that is used to define $(t,m,s)-$nets in base $b$. This connection will prove useful later on when we recall key results from \cite{wiart2021dependence} about the negative dependence properties of Faure and Halton sequences.  More precisely, Faure sequences in base $b$ and Halton sequences based on a vector $\BFb=(b_1,\ldots,b_s)$ of co-prime bases both have the property of being perfectly equidistributed, as per the following definition.

\begin{defi}
We  say that $P_n$ with $n$ of the form $b^m$ with $m \ge 0$ for a single-base $b$ construction  (and of the form $n=b_1^{m_1}\ldots b_s^{m_s}$ with $m_1,\ldots,m_s \ge 0$ for a multi-base $\BFb$ construction), is  {\em $(k_1,\ldots,k_s)$-equidistributed in base $b$ (or base $\BFb$)} if every {\em elementary $(k_1,\ldots,k_s)-$interval} of the form
\begin{equation}
\label{eq:elemInt}
I_{\BFk}(\BFa) = \prod_{\ell=1}^s \left[\left. \frac{a_{\ell}}{b_{\ell}^{k_{\ell}}},\frac{a_{\ell}+1}{b_{\ell}^{k_{\ell}}} \right.\right)
\end{equation}
for $0 \le a_{\ell} < b_{\ell}^{k_{\ell}}$ contains exactly $nb_1^{-k_1}\cdots b_s^{-k_s}$ points from $P_n$, assuming $\BFk$ is such that $n \ge b_1^{k_1} \ldots b_s^{k_s}$, and letting $b_{\ell}=b$ for a single base construction. 

We say that a single-base $b$ point set $P_n$ has a {\em quality parameter $t$} if $P_n$ is $(k_1,\ldots,k_s)$-equidistributed for all $s$-dimensional vectors of non-negative integers $\BFk=(k_1,\ldots,k_s)$ such that $k_1 + \ldots + k_s \le m-t$. We then refer to $P_n$ as a $(t,m,s)$-net in base $b$. 

For a sequence in base $b$, its quality parameter is $t$ if for every $m \ge 0$ and every point set of the form $\BFu_{j},\ldots,\BFu_{j+b^m-1}$, where $j$ is of the form $j=vb^m+1$ for some $v \ge 0$, is a $(t,m,s)-$net.
\end{defi}

Going back to the counting numbers $M_b(\BFk;P_n)$ introduced earlier, we can now say that for a given partition induced by the vector $\BFk$, it adds up the number of pairs of points that are found in each of the $b^{|\BFk|}$ elementary $\BFk-$intervals. So for instance if we only have one or zero point per elementary $\BFk-$interval, then $M_b(\BFk;P_n)=0$. Intuitively speaking, for a given number of points $n$ and partition induced by $\BFk$, if each elementary $\BFk-$interval contains the same number of points, then $M_b(\BFk;P_n)$ will be minimized. If all $n$ points are in one of the elementary $\BFk-$interval, then $M_b(k;P_n) = n(n-1)$ is maximal. If $P_n$ is a random point set, then for any given point, the expected number of distinct points from $P_n$ that are in the same elementary $\BFk-$interval is $(n-1)/b^{|\BFk|}$.
%
This leads us to the definition of the $C_b(\BFk;P_n)$ values, which play a key role in our analysis below. All results stated below are proved in \cite{wiart2021dependence}. The multi-base version is explored after.

\begin{defi}
\label{def:Cbk}
Let $P_n$ be a set of $n$ points in $[0,1)^s$ and $b \ge 2$ be an integer. Let $C_b(\BFk;P_n)$ be defined as
\[
C_b(\BFk;P_n) = \frac{b^{|\BFk|} M_b(\BFk;P_n)}{n(n-1)}.
\]
\end{defi}

It is easy to see that $C_b(\BFk;P_n) = 1$ when $\BFk=\mathbf{0}$.  
In the special case of a digital $(0,m,s)-$ net, we have the following result.
 
\begin{lemma}
If $P_n$ is a $(0,m,s)$-net in base $b$, then
\[
C_b(\BFk;P_n)= \frac{b^{|\BFk|}(\max(b^{m-{|\BFk|}}-1,0))}{b^m-1}.
\]
\end{lemma}

The $C_b(\BFk;P_n)$ values  have a connection with the concept of $\BFk-$equidistribution, as shown in the following lemma, also from \cite{wiart2021dependence}.

\begin{lemma} 
\label{lem:formCkiffequid}
Let $\BFk \in \mathbb{N}^s$ be such that $|\BFk| \le m$. A set $P_n \subseteq [0,1)^s$ with $n=b^m$ points is $\BFk-$equidistributed if and only if $C_b(\BFk;P_n)= b^{|\BFk|}(b^{m-|\BFk|} -1)/(b^m-1)$.
\end{lemma}

We also note that the value $C_b(\BFk;P_n)$ can be computed for any point set $P_n$ and base $b \ge 2$, and leads us to the introduction of  the concept of complete quasi-equidistribution in base $b$ from \cite{wiart2021dependence}.

\begin{defi}
\label{DefCqe}
Let $P_n$ be a point set of size $n$ in $[0,1)^s$ and $b \ge 2$ be a base. Let $\BFk = (k_1,\ldots,k_s) \in \N^s$. Then we say $P_n$ is {\em $\BFk-$quasi-equidistributed in base $b$} if $C_b(\BFk;P_n) \le 1$. If $C_b(\BFk;P_n) \le 1$ for all $\BFk \in \N^s$ then we say $P_n$ is {\em completely quasi-equidistributed (c.q.e) in base $b$}.
\end{defi}

Note that there are only finitely many values of $\BFk \in \N^s$ for which we need to compute $C_b(\BFk;P_n)$ in order to verify if $P_n$ is c.q.e.\!\,{:} see \cite{wiart2021dependence}. 

It is clear that a $(0,m,s)-$net in base $b$ is c.q.e (in base $b$), since by definition it is $\BFk-$equidistributed  for all $\BFk$ such that $|\BFk| \le m$. In  \cite{wiart2021dependence} it is also shown that the  first $n$ points of a $(0,s)-$sequence in base $b$ form a c.q.e.\ point set. 

To analyze Halton sequences we need a version of $C_b(\BFk;P_n)$ defined over a multi-base vector $\BFb$, as follows (see \cite{donglemieux22}).

\begin{defi}
\label{def:CbkHalton}
Let $P_n$ be a set of $n$ points in $[0,1)^s$ and $\BFb=(b_1,\ldots,b_s)$ with $b_j \ge 2$ for $j=1,\ldots,s$. Let $C_{\BFb}(\BFk;P_n)$ be defined as
\[
C_{\BFb}(\BFk;P_n) = \frac{{|\BFb|}^{|\BFk|} M_b(\BFk;P_n)}{n(n-1)}.
\]
where ${|\BFb|}^{|\BFk|} = \prod_{j=1}^s b_j^{k_j}$.
\end{defi}

In \cite{donglemieux22} it is shown that if $P_n$ is based on $n$ consecutive points from the Halton sequence or a generalized Halton sequence (which will be defined in the next section), then $C_{\BFb}(\BFk;P_n) \le 1$
for all $\BFk$. In other words, if we extend the concept of c.q.e.\ to constructions based on a multi-base vector $\BFb$, then the result from \cite{donglemieux22} shows these Halton point sets are c.q.e.



Now that we have introduced the key quantity $C_b(\BFk;P_n) $, we wish to explain how it relates back to negative dependence concepts. As shown in \cite{wiart2021dependence} and stated below in Theorem \ref{cor:NLODiffCle1}, a point set that is c.q.e.\ has a certain form of dependence captured by the concept of {\em negative lower orthant dependence}. In order to define this concept, we need to introduce a function denoted $H(\BFx,\BFy;\tilde{P}_n)$ defined over $[0,1]^{2s}$ for a 
sampling scheme  $\tilde{P}_n = \{\BFU_1,\ldots,\BFU_n\}$, 
where we assume each $\BFU_i$ is uniformly distributed over $[0,1)^s$ with a possible dependence structure between the $\BFU_i$'s. We then have:

\begin{equation}
\label{eq:puvdims}
H(\BFx,\BFy;\tilde{P}_n) := \frac{2}{n(n-1)}\sum_{i =1}^{n-1}\sum_{j > i} P(\BFU_i \le \BFx,\BFU_j \le \BFy),
\end{equation}
where $\BFU \le \BFx$ means $U_j \le x_j$ for all $j=1,\ldots,s$.

We can think of $H(\BFx,\BFy;\tilde{P}_n)$ as the joint distribution function of a pair of (distinct) points $(\BFU_I,\BFU_J)$ randomly chosen in $\tilde{P}_n$. (Here, we use capital letters for the indices $I$ and $
J$ to make it clear the points are randomly selected.) 
%

\begin{defi}
A sampling scheme $\tilde{P}_n$ is said to be an {\em NLOD sampling scheme} if $H(\BFx,\BFy;\tilde{P}_n) \le \prod_{\ell=1}^s x_{\ell}y_{\ell}$ for all $0 \le x_{\ell},y_{\ell} \le 1$.
\end{defi}

Intuitively speaking, the negative dependence that holds for an NLOD sampling scheme $\tilde{P}_n$ is desirable  because it implies the points are less likely to be clustered together, as they instead tend to repel each other, thus ensuring the sampling space is well covered by the points in $\tilde{P}_n$.

We can also look at how this negative dependence would affect the quality of an estimator for 
\begin{equation}
    \label{eq:muf}
\mu(f)=\int_{[0,1)^s} f(\BFx)d\BFx,
\end{equation}
based on the sampling scheme 
$\tilde{P}_n = \{\BFU_1,\ldots,\BFU_n\}$ 
via the  unbiased estimator 
\[
\hat{\mu}_n = \frac{1}{n} \sum_{i=1}^n f(\BFU_i).
\]

As discussed in  \cite{Lem17}, we have that
\[
{\rm Var}(\hat{\mu}_n) = \frac{\sigma^2}{n} + \frac{n-1}{n} {\rm Cov}(f(\BFU_I),f(\BFU_J)),
\]
where $\sigma^2 = {\rm Var}(f(\BFU))$ and ${\rm Cov}(f(\BFU_I),f(\BFU_J))$ represents the covariance term between the value of the integrand $f$ at two distinct, randomly chosen points $\BFU_I$ and $\BFU_J$ from $\tilde{P}_n$.
This covariance term differentiates the variance 
of $\hat{\mu}_n$---when $\tilde{P}_n$ is a dependent sampling scheme---from that of a Monte Carlo estimator with the same number  of points $n$.

This covariance can be written as 
\begin{equation}
\label{eq:covaspdf}
\sigma_{I,J} := {\rm Cov}(f(\BFU_I),f(\BFU_J)) = \int_{[0,1]^{2s}} f(\BFx) f(\BFy) \psi(\BFx,\BFy)d\BFx d\BFy -
\int_{[0,1]^{2s}} f(\BFx) f(\BFy)d\BFx d\BFy.
\end{equation}
where $\psi(\BFx,\BFy)$ is the joint pdf of $(\BFU_I,\BFU_J)$ evaluated at $(\BFx,\BFy)$. \dg{(We set $\psi(\BFx,\BFy)=0$ if one of the coordinates of $\BFx$ or $\BFy$ is equal to 1.)} In particular, this means we can also write

\begin{equation}
    \label{eq:HxyasInteg}
H(\BFx,\BFy;\tilde{P}_n) =  \int_{[\mathbf{0},\BFx) \times [\mathbf{0},\BFy)} \psi(\BFu,\BFv)d\BFu d\BFv.
\end{equation}
  Note that formally speaking, and as discussed in \cite{wiart2021dependence}, the definition of $H(\BFx,\BFy;\tilde{P}_n)$ given in \eqref{eq:puvdims} should lead to a closed  integration domain in \eqref{eq:HxyasInteg}. The reason why we instead integrate over a half-open interval is because it aligns better with the properties of $\psi(\BFx,\BFy)$, and is a convention we will follow throughout this paper. It is a valid approach because the boundary has measure 0, and thus the integral is unchanged whether we use a half-open interval or a closed one.

In this paper, we omit to say more on the joint pdf $\psi(\BFx,\BFy)$ corresponding to a base $b-$digitally scrambled point set  as it is not needed to present the results that connect the $C_b(\BFk;P_n)$ values to the NLOD property.

The next result, again from \cite{wiart2021dependence}, gives a necessary and sufficient condition for a digitally scrambled point set to be NLOD. The condition is based on the c.q.e.\ concept introduced in Definition \ref{DefCqe}, which holds when all $C_b(\BFk;P_n)$ values are no larger than 1.

In what follows, the term base-$b$ digital scrambling refers to either the method proposed by Owen in \cite{owen1995randomly} or the ``random linear scrambling'' proposed by Matou\v{s}ek \cite{matousek1998thel2}. The essential property that these scrambling methods have compared to others is that the corresponding joint pdf $\psi(\BFy,\BFy)$ is constant over each region $D_{\BFi}^s$, where $D_{\BFi}^s= \{(\BFx,\BFy) \in [0,1)^{2s}: \boldsymbol{\gamma}_b^s(\BFx,\BFy) = \BFi\}.$ 
These two scrambling approaches can also be applied to Halton sequences, as discussed in \cite{donglemieux22}.

\begin{thm}
\label{cor:NLODiffCle1}
Let $P_n$ be a deterministic point set of size $n$  in $[0,1)^s$ and $b \ge 2$ be an integer. Assume $P_n$ is such that the $j$th coordinate of the points are all distinct. Let ${}_b\tilde{P}_n$ be the sampling scheme obtained by applying a base $b-$digital scramble to $P_n$. Then ${}_b\tilde{P}_n$ is NLOD if and only if $P_n$ is c.q.e.
\end{thm}

In the special case of scrambled digital nets, the result can be stated more precisely \cite{wiart2021dependence}:

\begin{thm}\label{MainTheorem}
Let ${}_b\tilde{P}_n =\{\BFU_1,\ldots,\BFU_n\}$ be a scrambled digital $(t,m,s)$-net in base $b$, with $P_n$ such that its one-dimensional projections are digital $(0,m,1)$-nets. Then ${}_b\tilde{P}_n$ is an NLOD sampling scheme if and only if $t=0$.
\end{thm}

As a corollary to Theorem \ref{cor:NLODiffCle1}, in the case of Halton sequences we have the following result from \cite{donglemieux22}:

\begin{cor}
A scrambled Halton point set is an NLOD sampling scheme. 
\end{cor}

We now explain how the $C_b(\BFk;P_n)$ values can be used as a common framework to assess deterministic and random scramblings.

On one hand, for point sets that are c.q.e.---such as those arising from the Faure and Halton sequences and their generalizations---a base $b$-scrambling outputs a randomized point set $\tilde{P}_n$ that is NLOD, which for the purpose of this paper we assume means is amenable to improvement over the Monte Carlo method for some functions. More generally speaking, we view this property as implying good uniformity properties \cite{unanchored2022, wiart2021dependence}.
We note that for point sets that are not c.q.e., such as the Sobol' sequences, a base-$b$ scrambling is unable to yield the NLOD property, which is consistent with observations made by others that scrambling does not always manage to ``fix'' bad projections of the Sobol' sequence \cite[p.497]{owensobol}, \cite{joyboyletan}.

Now if we instead want to use the $C_b(\BFk;P_n)$ values  to assess the quality of a deterministic point set $P_n$, we argue that only considering $C_b(\BFk;P_n)$ or $C_{\BFb}(\BFk;P_n)$ is not enough, as a small value of $C_b(k;P_n)$ simply means that if we scramble in base $b$, we will obtain a good quality point set. If we want to analyze the quality of the point set without the help it may get from scrambling, we should ensure that regardless of the base in which we may scramble, the point set would be of good quality. This is because while scrambling can improve the quality of a point set, it also cannot make it worse, in some sense. Hence for a point set constructed in base $b>2$ (or in multiple bases such as the Halton sequence), if $C_{b'}(\BFk;P_n)$ is small for $b'\neq b$, then it suggests that the particular form of scrambling used is not important, which we argue is more likely to hold if the point set is already well distributed.
For this reason, for point sets constructed in bases larger than 2, we propose to use quality measures based on $C_2(\BFk;P_n)$ to assess their quality. Details are given in Section \ref{sec:compcbk}. 


\section{Good deterministic scramblings}
\label{sec:detscr}

Many ideas have been proposed to improve the quality of the Faure and Halton sequences which are the focus of this paper. Before going any further, we start by recalling how the original constructions are defined and then we describe the type of deterministic scramblings that lead to various generalizations. We then explain how permutations proposed in \cite{faure1992good} can be used to define both generalized Halton and generalized Faure sequences.

We follow the notation and framework of \cite{faure2009generalized} also used in \cite{donglemieux22}.
The building block for the Halton sequence is  the {\em van der Corput sequence in base $b$}, denoted $S_b$, which has its $n^{\text{th}}$ term ($n \ge 1$) defined as
\begin{equation}
\label{eq:vdc}
S_b(n) = \sum_{r=0}^{\infty} \frac{a_r(n)}{b^{r+1}},
\end{equation}
where $a_r(n)$ is the $r^{\text{th}}$ digit of the $b$-adic expansion of
$n-1=\displaystyle\sum_{r=0}^\infty a_r(n)\ b^r$.

The {\em Halton sequence} is an $s$-dimensional sequence  $\BFu_1,\BFu_2,\ldots$ in $[0,1)^s$ defined as
\begin{equation}
\label{eq:defHalton}
\BFu_n = (S_{b_1}(n),\ldots,S_{b_s}(n)),
\end{equation}
where the $b_j$'s, for $j=1,\ldots,s$, are pairwise coprime. 
That is, the $j^{\text{th}}$ coordinate is defined using $S_{b_j}$, the van der Corput sequence in base $b_j$. These $b_j$'s are typically chosen as the first $s$ prime numbers. 

A {\em generalized van der Corput sequence} \cite{faure1992good} is obtained by choosing a sequence $\Sigma = (\sigma_r)_{r \ge 0}$ of permutations of $Z_b=\{0,1,\ldots,b-1\}$. Then, the $n^{\text{th}}$ term of the sequence is defined as
\begin{equation}
\label{eq:genVDC}
S_b^{\Sigma}(n) = \sum_{r=0}^\infty \frac{\sigma_r\big(a_r(n)\big)}{b^{r+1}}.
\end{equation}
If the same permutation $\sigma$ is used for all digits (i.e., if $\sigma_r=\sigma$ for all $r \ge 0$), then we use the notation $S_b^{\sigma}$ to denote  $S_b^{\Sigma}$.
The van der Corput sequence in base $b$ defined in (\ref{eq:vdc}) is obtained by taking $\sigma_r = I$ for all $r \ge 0$, where $I$ stands for the identity permutation over $Z_b$.

A {\em generalized Halton sequence} is obtained by choosing $s$ sequences of permutations
$\Sigma_j = (\sigma_{j,r})_{r \ge 0}$,
$j=1,\ldots,s$, and defining the $n^{\text{th}}$ point
as
\begin{equation}
\label{def:GHS}
\BFu_n = (S_{b_1}^{\Sigma_1}(n),\ldots,S_{b_s}^{\Sigma_s}(n)),\,\, n \ge 1.
\end{equation}
Examples of generalized Halton sequences can be found in \cite{faure2009generalized,donglemieux22} and the references therein.

Another way to generalize the van der Corput sequence is to apply well-chosen linear transformations to the digits $a_r(n)$ before multiplying them by $b^{-(r+1)}$ in \eqref{eq:vdc}. More precisely, let $C$ be an $\infty \times \infty$ matrix with elements in $\mathbb{Z}_b$, where we assume $b$ is prime. 
Let $C_{r,\ell}$ be the element on the $r^{\text{th}}$ row and $\ell^{\text{th}}$ column of $C$.
The $n^{\text{th}}$ term is then defined as
\begin{equation}
\label{eq:linVDC}
S_b^{C}(n) = \sum_{r=0}^\infty 
\sum_{\ell=0}^{\infty} C_{r,\ell} a_{\ell}(n) {b^{-(r+1)}}.
\end{equation}
Using different transformations $C_j$ on a given van der Corput sequence in a fixed base $b$ to obtain the $j^{\text{th}}$ coordinate of a point in $[0,1)^s$ breaks the clearly wrong pattern of  an $s$-dimensional sequence that would otherwise have all its points along the main diagonal in $[0,1)^s$.  The Faure sequence is obtained in a prime base $b \ge s$ and using successive powers of the Pascal matrices to define the matrices $C_j$ \cite{faure1982}. A {\em generalized Faure sequence} is obtained by pre-multiplying these matrices by a non-singular lower triangular matrix \cite{rTEZ95a}. Examples of generalized Faure sequences can be found in \cite{rTEZ94b,qFAU09a}.


In this paper we explore the use of the permutations proposed in \cite{faure1992good} to construct both generalized Halton and Faure sequences. Before we explain our proposed constructions, we first provide some background on these permutations.


In fact, the paper \cite{faure1992good} is a by-product of the seminal article \cite{Faure81}, where an extensive study of generalized van der Corput sequences is developed by means of so-called $\varphi$-functions (see Section 2.1 p. 49). This paper is written in French but a condensed English version exists within the survey paper \cite{FKP}.
In the latter, in Subsection 2.4, p.776,  $\varphi$-functions are introduced in the so-called elementary descent method at the origin of exact formulas for the star-discrepancy. Then, Algorithm 3.1 of \cite{faure1992good} is evoked together with the concept of Intrication (see \cite[p. 801, Definition 8, Lemma 5]{FKP}) and leads to the result provided in Theorem 38 of \cite{FKP}, which  shows that a generalized van der Corput sequence in base $b$ obtained by applying this permutation achieve a bound of $1/\log 2$ for its normalized star-discrepancy (i.e., the star-discrepancy is bounded by $\log n/(n \log 2)$). 

These proposed permutations are especially useful because they can be obtained via a recursive process (over the bases $b$) consisting of a descent method, which is illustrated using checkerboards in \cite[Sec.\ 3.1]{FKP}. As can be seen there, even bases (and thus checkerboards with an even dimension) can be intricated together to create a larger point set. To get a permutation for an odd base, a point needs to be inserted in the middle of an even-sized checkerboard. 
The process begins with the two first points (i.e., base 2) and grows. This construction is useful because of its simplicity and step-by-step induction. But as mentioned in \cite[Subsection 3.4]{FKP}, there are many other ways to get good permutations, which could lead to future  investigations.

Algorithm \ref{algo:faure1992} summarizes
the steps for generating a permutation $\pi_b$ for any integer $b$ (in fact, the recursive algorithm generates all permutations $\pi_j$ for $j \leq b$).

\begin{algorithm}[ht]
INPUT:b\;
OUTPUT: list of permutations ($\pi_2$,...,$\pi_b$)\;
j = 2\;
$\pi_j$ = (0,1)\;
\While{$j \leq b$:}{
j = j + 1\;
\If{j is even:}{
$\pi_j$ = ($2\pi_{j/2}$, $2\pi_{j/2}+1$)\;
}
\If{j is odd:}{
temp = $\pi_{j-1}$\;
k = $\frac{j-1}{2}$\;
\For{$\ell=1$ to $j-1$:}{
\If{$\pi_j[\ell] \ge k$:}{$\pi_j[\ell] = \pi_j[\ell]+1$}}
\For{$\ell=k+1$ to $j$:}{$\pi_j[\ell+1]=\pi_j[\ell]$}
$\pi_j[k+1]=k$\;
}
}
\Return ($\pi_2$,...,$\pi_b$)
\caption{Algorithm to generate permutations from \cite{faure1992good}}
\label{algo:faure1992}
\end{algorithm}

While these permutations have been proposed to improve the discrepancy of one-dimensional van der Corput sequences, they can be (and have been) used to create generalized Halton sequences \cite{okten2009generalized}, which is what we propose to analyze via dependence concepts in the next section.


Next we propose to use these permutations to create generalized Faure sequences using diagonal NLT matrices
each based on a factor $f_j$, $j=1,\ldots,s$. 
Clearly, the factors $f_j$ should be non-zero, so we cannot simply use 
$f_j = \pi_b[j]$. We propose two ways to address this, 
and in both cases, we only obtain $b-1$ factors, and hence can only define a $(b-1)-$dimensional sequence based on this method (rather than a full $b-$dimensional sequence). 

With the first method, we observe that all permutations from \cite[Subsection 3.4]{faure1992good} are such that $\pi_b[0]=0$, so we simply omit the first term of the permutation. With the second method, we add an ``offset term'' $m = \lfloor b/2 \rfloor$ (modulo $b$) to each term of the permutation, which based on the algorithm used to generate $\pi_b$, is such that we now have $\pi_b[m]=0$. That is, the addition of the offset term has the effect of placing the 0 in the middle of the permutation vector. More precisely, the two different methods for using the permutations from \cite{faure1992good} to define factors for the Faure and Halton sequences are: 

\begin{enumerate}
    \item Let  $f_{j,1}=\pi_b[j+1]$ for $0 \le j \le b-2 $.
    \item Let $f_{j,2} = (\pi_b[j] + m) \bmod b$ for $j=0,\ldots,m-1$ and $f_j = (\pi_b[j+1] + m) \bmod b$ for $j=m,\ldots,b-2$
    \hchristiane{Should prove that this moves the 0 in the last spot}
\end{enumerate}

In our numerical results presented in Sections \ref{sec:compcbk} and \ref{sec:num}, we use the second method above to reorder the permutations from \cite{faure1992good} and then apply them to the Faure and Halton sequences. 
We refer to that approach as the ``Faure 1992 Offset" factors. In the remainder  of the paper, in tables and histograms, for short we use the abbreviation ``F92" factors to refer to the first method and ``F Offset" factors to refer to the second one.

\section{Comparisons based on negative dependence criterion}
\label{sec:compcbk}

As explained at the end of Section \ref{sec:negdep}, 
here we propose to compare different point sets using the quantity $C_2(\BFk;P_n)$ for different vectors $\BFk$ corresponding to partitions of the unit hypercube $[0,1)^s$ that are of particular interest.

Define $c(b,{\cal K};P_n) = \max_{\BFk \in {\cal K}} C_b(\BFk;P_n)$, where ${\cal K} \subset \mathbb{N}^s$. For a given dimension $s$, we are interested in a special class of subsets ${\cal K}$ defined by two parameters $(d,w)$ where $2 \le d \le s$ and $w \ge 1$, in the following way:
\[
{\cal K}_{d,w,s} = \{\BFk \in \mathbb{N}^s: 2 \le \sum_{j=1}^d \mathbf{1}_{k_j>0} \le d, r({\bf k}) \le w \},
\]
where $r(\BFk) = \max \{1 \le j \le s: k_j>0\}
- \min \{1 \le j \le s: k_j>0\}$ can be thought as the \emph{range} of indices where a non-zero component of $\BFk$ can be found.
Hence $d$ refers to the largest number of non-zero components allowed for $\BFk$ to be included in ${\cal K}_{d,w,s}$  and $w$ is in some sense a window size, which limits the range of the indices $j$ where a non-zero $k_j$ is found.

For instance, if $s=4$, $d=2$ and $w=2$, then
\begin{align*}
{\cal K}_{d,w,s} =& \{(k,\ell,0,0),(k,0,\ell,0),(0,k,\ell,0),(0,k,0,\ell),
(0,0,k,\ell): k,\ell>1\}.
\end{align*}
That is, we have excluded all vectors $\BFk$ with only 1 non-zero component or with more than 2 non-zero components, and we also excluded vectors such as $(1,0,0,3)$ because its range is 3 which is larger than the largest allowed range of $w=2$.

As an alternative to the criterion $c(b,{\cal K};P_n)$ which returns the worst (largest) $C_b(\BFk;P_n)$ value, we also consider one based on the mean, defined as
$\bar{c}(b,{\cal K};P_n) = \frac{1}{|{\cal K}|}\sum_{\BFk \in {\cal K}} C_b(\BFk;P_n)$.

In Tables \ref{tab:critF1} and \ref{tab:critF3}, we compute these two criteria for point sets based on the (generalized) Faure sequence, where ``Faure 1992'' and ``F Offset'' respectively refer to the two methods for fixing the factors $f_j$ mentioned at the end of Section \ref{sec:detscr}, and ``Regular'' refers to the original construction from \cite{faure1982}. The tables differ in the samples size $n$ used for the point sets.

Tables \ref{tab:critH1} and \ref{tab:critH3} are the counterpart tables for the (generalized) Halton sequence.
Similarly, here ``Regular'' refers to the original construction from \cite{rHAL60a}, ``Faure 1992'' refers to the permutations obtained using the method from \cite{faure1992good}, as detailed in  Algorithm \ref{algo:faure1992}, while ``F Offset'' refers to the permutation obtained when we add $m=\lfloor b/2 \rfloor$ to each term of the permutation, as explained at the end of Section \ref{sec:detscr}.


\begin{table}[ht]
\begin{center}
\begin{tabular}{lllllll}
\hline
Faure Permutations   & $b$  & $s$  & $d$ & $n$    & $c(2,\mathcal{K}_{d,w=s,s},P_n)$ & $\bar{c}(2,\mathcal{K}_{d,w=s,s},P_n)$  \\
\hline
Regular      & 5  & 4  & 2               & 3125 & 0.99968  & 0.99968  \\
Faure 1992  & 5  & 4  & 2               & 3125 & 0.99968  & 0.99968  \\
F Offset & 5  & 4  & 2               & 3125 & 0.99968  & 0.99968  \\
Regular      & 13 & 12 & 2               & 2197 & 1.755691 & 1.422891 \\
Faure 1992  & 13 & 12 & 2               & 2197 & 1.782858 & 1.162811 \\
F Offset & 13 & 12 & 2               & 2197 & 1.782858 & 1.136818 \\
Regular      & 53 & 52 & 2               & 2809 & 6.5223   & 1.584402 \\
Faure 1992  & 53 & 52 & 2               & 2809 & 14.88912 & 2.019255 \\
F Offset & 53 & 52 & 2               & 2809 & 14.88912 & 1.892771 \\
\hline
\end{tabular}
\caption{Values of criteria based on $C_2(\BFk;P_n)$ for point sets obtained from (generalized) Faure Sequences.}
\label{tab:critF1}
\end{center}
\end{table}

\begin{table}[ht]
\begin{center}
\begin{tabular}{lllllll}
\hline
Halton Permutations   & $s$ & $d$ & $n$ & $c(2,\mathcal{K}_{d,w=s,s},P_n)$ & $\bar{c}(2,\mathcal{K}_{d,w=s,s},P_n)$  \\
\hline
Regular & 4 & 2 & 3125 & 0.999685 & 0.999683 \\
Faure 1992  & 4 & 2 & 3125 & 0.999682 & 0.999682 \\
Offset  & 4 & 2 & 3125 & 0.999681 & 0.99968 \\
DL & 4 & 2 & 3125 & 0.999681 & 0.999681 \\
FL  & 4 & 2 & 3125 & 0.999682 & 0.999682 \\
Regular & 12 & 2 & 2197 & 4.036052 & 1.892203 \\
Faure 1992  & 12 & 2 & 2197 & 1.819365 & 1.071718 \\
Offset  & 12 & 2 & 2197 & 1.635136 & 1.053001 \\
DL  & 12 & 2 & 2197 & 3.531758 & 1.319083 \\
FL  & 12 & 2 & 2197 & 2.029063 & 1.18698 \\
Regular & 52 & 2 & 2809 & 13.65113 & 6.534291 \\
Faure 1992 & 52 & 2 & 2809 & 4.204183 & 1.21624 \\
Offset & 52 & 2 & 2809 & 4.179257 & 1.197399 \\
DL & 52 & 2 & 2809 & 11.092064 & 1.862851 \\
FL & 52 & 2 & 2809 & 8.539747 & 1.467202 \\
\hline
\end{tabular}
\caption{Values of criteria based on $C_2(\BFk;P_n)$ for point sets obtained from (generalized) Halton Sequences.}
\label{tab:critH1}
\end{center}
\end{table}

\begin{table}[ht]
\begin{center}
\begin{tabular}{lllllll}
\hline
Faure Permutations   & $b$  & $s$  & $d$ & $n$    & $c(2,\mathcal{K}_{d,w=s,s},P_n)$ &  $\bar{c}(2,\mathcal{K}_{d,w=s,s},P_n)$ \\
\hline
Regular      & 5  & 4  & 2               & 5000 & 1.19561576 & 1.0977079 \\
Faure 1992  & 5  & 4  & 2               & 5000 & 3.35611442 & 1.4793424 \\
F Offset & 5  & 4  & 2               & 5000 & 3.35611442 & 1.5871413 \\
Regular      & 13 & 12 & 2               & 5000 & 7.71906317 & 3.4452165 \\
Faure 1992  & 13 & 12 & 2               & 5000 & 9.54395039 & 1.9958566 \\
F Offset & 13 & 12 & 2               & 5000 & 7.63516031 & 2.1053137 \\
Regular      & 53 & 52 & 2               & 5000 & 111.422999 & 16.197021 \\
Faure 1992  & 53 & 52 & 2               & 5000 & 170.658418 & 4.89776   \\
F Offset & 53 & 52 & 2               & 5000 & 153.710041 & 4.696813  \\
\hline
\end{tabular}
\caption{$C_2$ values of point sets based on the Faure Sequence with $n = 5000$.}
\label{tab:critF3}
\end{center}
\end{table}

\begin{table}[ht]
\begin{center}
\begin{tabular}{lllllll}
\hline
Halton Permutations & $s$ & $d$ & $n$ & $c(2,\mathcal{K}_{d,w=s,s},P_n)$ & $\bar{c}(2,\mathcal{K}_{d,w=s,s},P_n)$ \\
\hline
Regular & 4 & 2 & 5000 & 0.99980252 & 0.99980183 \\
Faure 1992  & 4 & 2 & 5000 & 0.9998014 & 0.99980089 \\
F Offset & 4 & 2 & 5000 & 0.99979996 & 0.99979996 \\
DL  & 4 & 2 & 5000 & 0.99980012  &  0.99980004 \\
FL & 4 & 2 & 5000 & 0.9998014 & 0.99980076 \\
Regular & 12 & 2 & 5000 & 4.68872783 & 2.13327165 \\
Faure 1992  & 12 & 2 & 5000 & 1.47878792 & 1.05099389 \\
F Offset & 12 & 2 & 5000 & 1.40865037 & 1.03312189 \\
DL  & 12 & 2 & 5000 & 3.18306477 & 1.21785275 \\
FL & 12 & 2 & 5000 & 2.48300028 & 1.22527534\\
Regular & 52 & 2 & 5000 & 18.82439336 & 4.81624748 \\
Faure 1992  & 52 & 2 & 5000 & 4.73002376 & 1.24112783 \\
F Offset  & 52 & 2 & 5000 & 4.62842264 & 1.21837901 \\
DL  & 52 & 2 & 5000 & 10.95981116 &  1.75041329 \\
FL & 52 & 2 & 5000 & 6.94886113 & 1.54795824\\
\hline
\end{tabular}
\caption{$C_2$ values of point sets based on the Halton Sequence with $n = 5000$.}
\label{tab:critH3}
\end{center}
\end{table}

For the Faure sequence, we can see that even though the maximum value taken by $C_2(\BFk;P_n)$ is not smaller for the two generalized Faure sequences compared to the original Faure sequence, the average over all two-dimensional projections is smaller. 
This means that after multiplying the generating matrices by the chosen factors (Faure 1992 or F Offset), there are fewer "poor" two-dimensional projections, but the poorest projections are worse than prior to applying the factors. 
For the Halton sequence, applying permutations 
clearly improves the quality of the original construction. 
When comparing Tables \ref{tab:critF3} and \ref{tab:critH3}---which both use $n=5000$---the Halton sequence and its generalizations appear to be of better quality compared to their respective counterpart for the Faure sequence, with the generalized Halton sequences achieving values for  $\bar{c}(2,\mathcal{K}_{d,w=s,s},P_n)$ that are not much higher than 1. We think this may be happening because the permutations introduce a more substantial ``scrambling'' than simply multiplying each generating matrix by a fixed factor.




\section{Numerical experiments}
\label{sec:num}

In this section, 
we numerically investigate the fundamental question of whether it is best to randomize via random scrambling or to apply a well-chosen deterministic scrambling that can then be randomized using a simple digital shift. The latter are based on the constructions discussed in Section \ref{sec:detscr} and evaluated in Section \ref{sec:compcbk}.

\hgracia{I will reformat all the plots to make them look nicer/fix the legends}

To compare the different constructions we use the following test functions, where we assume the goal is to integrate them, i.e., compute $\mu(f)$ as defined in \eqref{eq:muf}:
\begin{enumerate}
    \item $h_0(\BFx) = \sum_{j=1}^s (e^{x_j}-e+1)$, function $g_0$ from \cite{owen2020dropping}, which integrates to 0 over the unit cube; 
    \item $h_1(\BFx) = (\sum_{j=1}^s x_j)^2$, function $g_1$ from \cite{owen2020dropping}, which integrates to $s/3+s(s-1)/4$ over the unit cube;
    \item Stochastic activity network (SAN):
    This is a 13-dimensional problem described in \cite{lemieux2009mcqmc}, page 99; however here we use 
a 12-dimensional version, with activity 10 removed and all other parameters the same.  A probability is estimated using naive Monte Carlo, which means the corresponding integrand is an indicator function.
    \item $g_2(\mathbf{x}) = \prod_{j=1}^s \left(1 + c(x_j - 0.5)\right)$, where $c$ is a constant, proposed in \cite{asotsky2003one}. This function has fairly low effective dimension and its ANOVA representation is given in \cite{asotsky2003one}. The true value of the integral is 1 over the unit cube. We consider two combinations of $s$ and $c$.  First, we use $s = 120, c = 0.1$, which has an effective dimension of 4 in the superposition sense with a threshold of 0.99; second, we use $s = 96, c = 0.25$, which has an effective dimension of 6 in the superposition sense with a threshold of 0.99.

\end{enumerate}

We compare estimators using either their mean-squared error (MSE) or variance, which are obtained as follows: when integrating a function $f$, we assume that for $v=1,\ldots,V$, $P_{n,v} = \{\BFx_{1,v},\ldots,\BFx_{n,v}\}$ is an $s$-dimensional point set that has been randomized, and that the $P_{n,v}$ for $v=1,\ldots,V$ are independent of each other (e.g., they have been randomized using independent shifts or independent scramblings). 
Then we either compute
\[
{\rm MSE} = \frac{1}{V} \sum_{v=1}^V \left(\hat{\mu}_{v} - \mu(f)\right)^2,
\]
where $\mu(f) = \int_{[0,1)^s} f(\BFx) d\BFx$ if the true value of the integrand is known, or
\[
{\rm Var} = \frac{1}{V-1} \sum_{v=1}^V \left(\hat{\mu}_{v} - \hat{\mu}\right)^2, 
\]
where $\hat{\mu} = \frac{1}{V} \sum_{v=1}^V \hat{\mu}_{v} $ is the sample mean over all $V$ independent randomizations if the true value of the integrand is not known. 
In both cases,
\[
\hat{\mu}_{v} = \frac{1}{n}\sum_{i=1}^n f(\BFx_{i,v}).
\]
That is, $\hat{\mu}_{v}$ is the estimator for $\mu$ obtained from the $v$th randomized point set $P_{n,v}$.
As noted above, for $f$ given by $h_0$, $h_1$, and $g_2$ we respectively have $\mu(f)= 0$, $\mu(f) = s/3+s(s-1)/4$ and $\mu(f) = 1$. 

The different constructions and scramblings are compared using two approaches. First in Section \ref{sec:convergence}
we plot the MSE or variance as a function of $n$ to see how quickly they converge to 0. Then in Section \ref{sec:hist}, we fix $n$ and plot a histogram of the MSE or Variance constructed from randomly scrambled point sets and look at where different deterministically scrambled point sets compare with this distribution, as approximated by the histogram.





\subsection{Convergence Results for the Faure and Halton sequences}
\label{sec:convergence}

In all the results of this section, the Faure sequence in constructed in base $b$ where $b$ is the smallest prime number $b$ such that $b \geq s$. 
Note that for this experiment, all values of $n$ plotted in the graph are an integer multiple of an integer power of this base $b$.

 

\begin{figure}[htb]
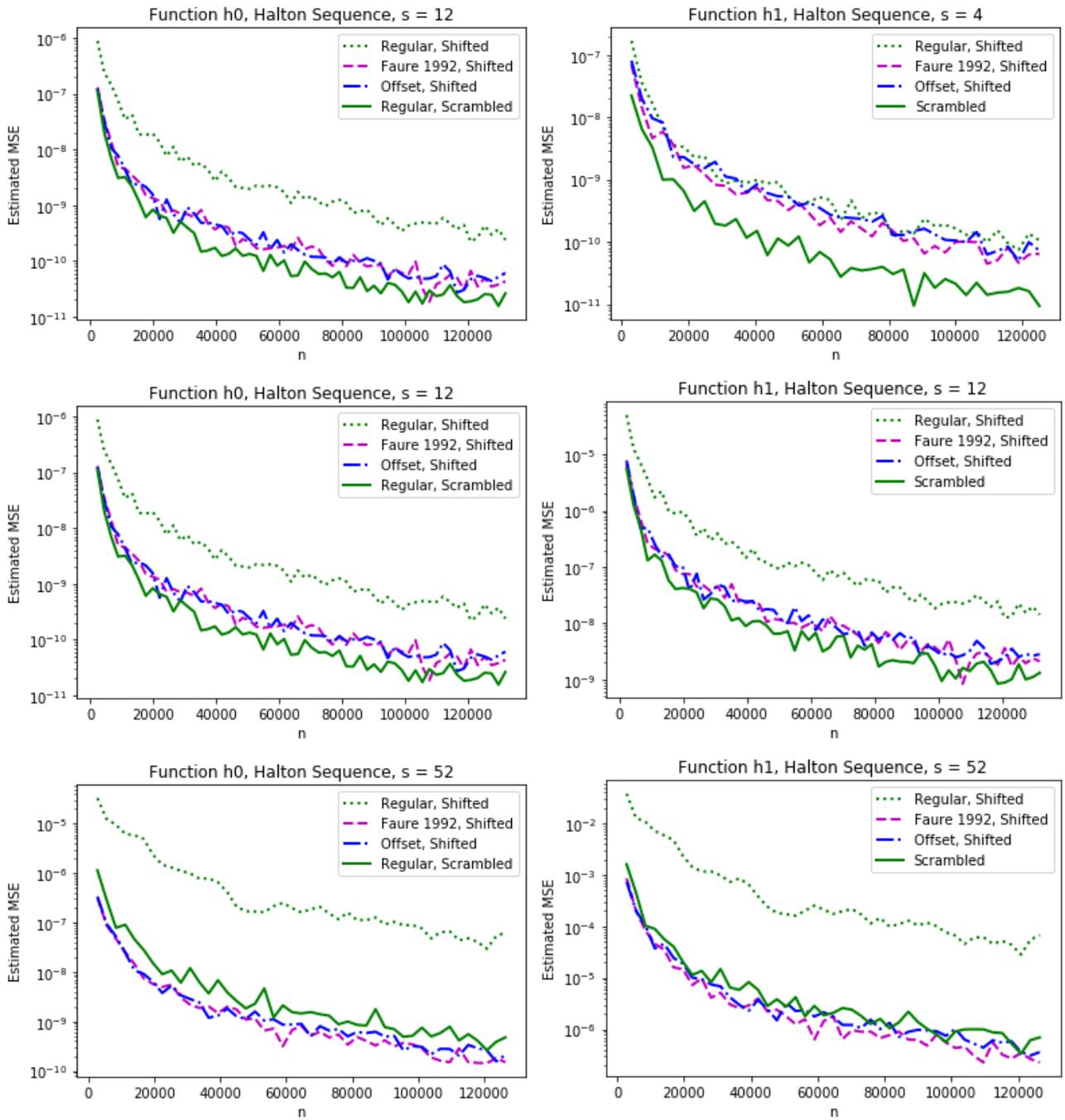

 \centering
 \includegraphics[width=0.5\textwidth]{haltonh0s12.png}\hfil
 \includegraphics[width=0.5\textwidth]{haltonh1s4.png}\hfil 
 \includegraphics[width=0.5\textwidth]{haltonh0s12.png}\hfil
 \includegraphics[width=0.5\textwidth]{haltonh1s12.png}\hfil 
 \includegraphics[width=0.5\textwidth]{haltonh0s52.png}\hfil
 \includegraphics[width=0.5\textwidth]{haltonh1s52.png}\hfil 
 \caption{Estimated MSEs test functions using the 4-, 12- and 52-dimensional Halton Sequence}
 \label{fig:HaltonTestFuncs}
 \end{figure}

What is plotted on Figures \ref{fig:ConvBase5} to \ref{fig:g2MSE} are the MSE for functions $h_0$, $h_1$, and $g$, and the variance for the SAN. In all cases the MSE or variance is estimated using $V=25$ randomizations.

For the Faure sequence, we compare the following constructions: 
\begin{enumerate}
    \item Faure sequence, randomized with a digital shift (``Regular, Shifted'');
    \item Generalized Faure sequence, using factors from Algorithm \ref{algo:faure1992}, then randomized with a digital shift (``Faure 1992, Shifted'');
    \item Generalized Faure sequence, using factors from Algorithm \ref{algo:faure1992} with an offset term added, then randomized with a digital shift (``Offset, Shifted'');
    \item Faure sequence, randomized with Owen's scrambling (``Regular, Scrambled'').
\end{enumerate}

For the Halton sequence, we compare the following constructions: 
\begin{enumerate}
    \item Halton sequence, randomized with a digital shift, (``Regular, Shifted'');
    \item Generalized Halton sequence, using permutations
    from Algorithm \ref{algo:faure1992}, then randomized with a digital shift (``Faure 1992, Shifted'');
    \item Generalized Halton sequence, using permutations from Algorithm \ref{algo:faure1992} with an offset term added, then randomized with a digital shift (``Offset, Shifted'');
    \item Halton sequence, randomized with random linear scrambling (``Scrambled'').
\end{enumerate}

\statement{
\begin{figure}[htb]
\centering
\includegraphics[width=0.5\textwidth]{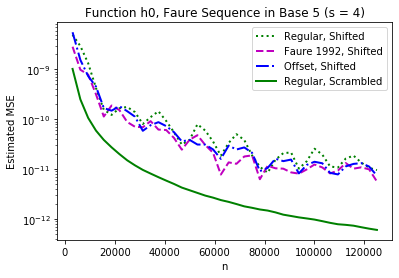}\hfil
\includegraphics[width=0.5\textwidth]{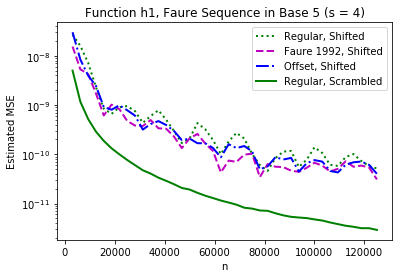}\hfil
\includegraphics[width=0.5\textwidth]{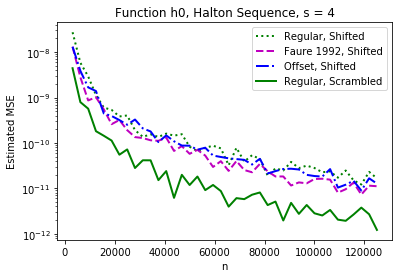}\hfil
\includegraphics[width=0.5\textwidth]{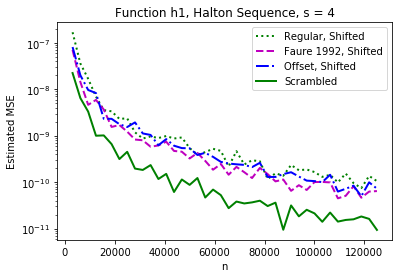}\hfil
\caption{Estimated MSEs of test functions at $n \in \{3125m,1 \le m \le 40\}$, using the 4-dimensional Faure Sequence constructed in base 5, using factors $[3,2,1,4]$ (Faure 1992) and $[3,1,4,2]$ (Offset) and the Halton sequence with the corresponding permutations}
\label{fig:ConvBase5}
\end{figure}

\begin{figure}[htb]
\centering
\includegraphics[width=0.5\textwidth]{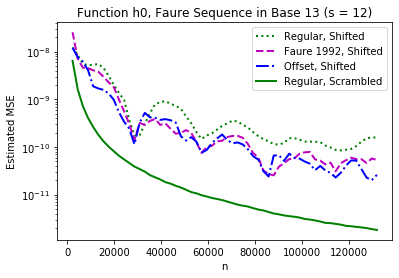}\hfil
\includegraphics[width=0.5\textwidth]{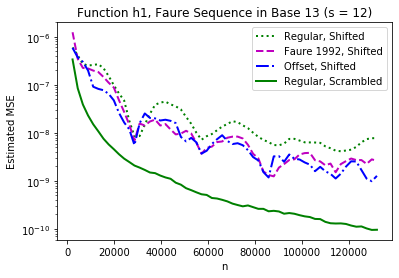}\hfil
\includegraphics[width=0.5\textwidth]{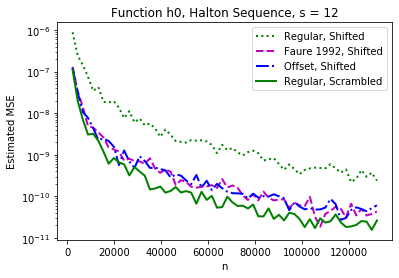}\hfil
\includegraphics[width=0.5\textwidth]{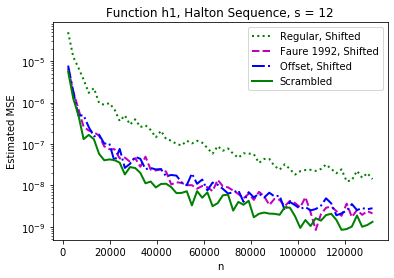}\hfil
\includegraphics[width=0.5\textwidth]{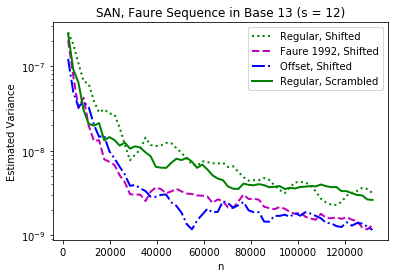}\hfil
\includegraphics[width=0.5\textwidth]{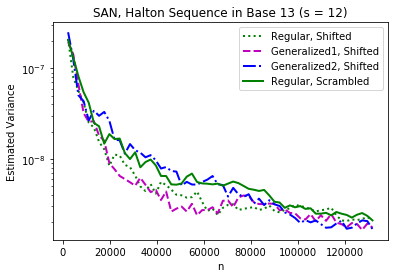}\hfil
\caption{Estimated MSEs and Variances at $n \in \{2197m,1 \le m \le 60\}$ of test functions using the 12-dimensional Faure Sequence constructed in base 13, using factors $[4,  9,  2,  7, 11,  6,  1,  5, 10,  3,  8, 12]$ (Faure 1992) and $[ 7, 11,  3,  9,  1,  5,  8, 12,  4, 10,  2,  6]$ (Offset) and the Halton sequence with the corresponding permutations.}
\label{fig:ConvBase13}
\end{figure}
\hgracia{I reran the 12-dim SAN problem (data saved resFaureb13\_SAN\_20220108.pickle for if I need to regenerate this plot)}

\begin{figure}[htb]
\centering
\includegraphics[width=0.5\textwidth]{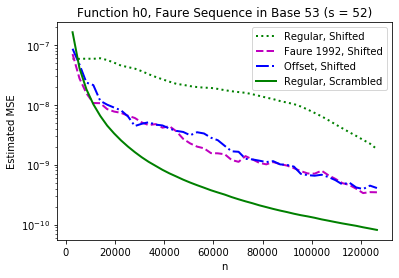}\hfil
\includegraphics[width=0.5\textwidth]{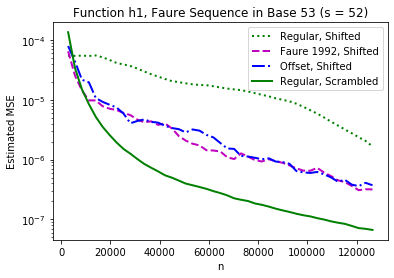}\hfil 
\includegraphics[width=0.5\textwidth]{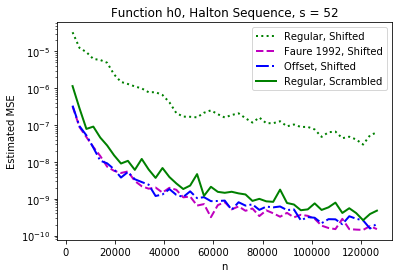}\hfil
\includegraphics[width=0.5\textwidth]{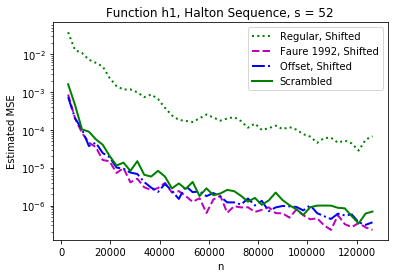}\hfil

\caption{Estimated MSEs of test functions at $n \in \{2809m, 1\le m \le 45\}$, using the 52-dimensional Faure Sequence constructed in base 53, using factors $[16, 37, 8, 29, 45, 24, 4, 20, 41, 12, 33, 49, 2, 18, 39, 10, 31, 47, 27, 6, 22, 43, 14, 35, 51, 26, 1, 17, 38, 9, 30, 46, 25, \\ 
5, 21, 42, 13, 34, 50, 3, 19, 40, 11, 32, 48, 28, 7, 23, 44, 15, 36, 52]$ (Faure 1992) and $[27, 43, 11, 35, 3, 19, 51, 31, 47, \\
15, 39, 7, 23, 29, 45, 13, 37, 5, 21, 1, 33, 49, 17, 41, 9, 25, 28, 44, 12, 36, 4, 20, 52, 32, 48, 16, 40, 8, 24, 30, 46, 14, 38, 6, \\
22, 2, 34, 50, 18, 42, 10, 26]$(Offset) and the Halton sequence with the corresponding permutations.}
\label{fig:ConvBase53}
\end{figure}
}

\statement{
\begin{figure}[htb]
\centering 
  \includegraphics[width=0.5\textwidth]{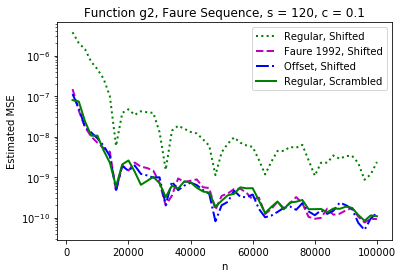}\hfil
  \includegraphics[width=0.5\textwidth]{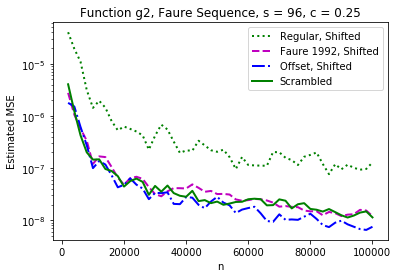}\hfil
    \includegraphics[width=0.5\textwidth]{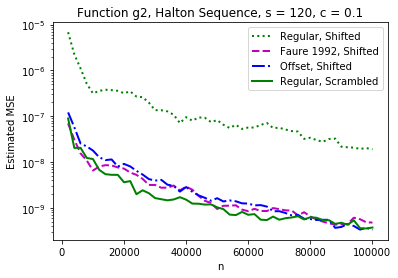}\hfil
  \includegraphics[width=0.5\textwidth]{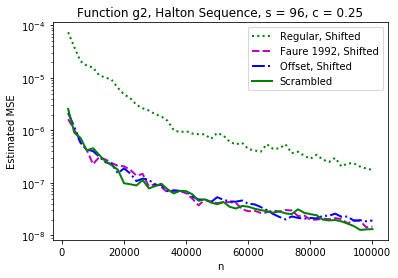}\hfil
\caption{Estimated MSEs and Convergence Rate at $n = \in \{2000m, 1\le m \le 50\}$, when integrating Test Function $g$ using Halton and Faure Sequences.}
\label{fig:g2MSE}
\end{figure}
}

The majority of the results show that scrambling is superior to generalizing then randomizing with a digital shift. 
For the Faure sequence, when $n$ is an integer multiple of a larger power of the constructing base $b$, we can see an improvement in performance for the constructions randomized with a simple digital shift. 
\hchristiane{by shifted we mean using F92 or F Offset with a digital shift, correct? Maybe say ``the generalized constructions based on F92 or F Offset randomized with a digital shift''?}
For the Halton sequence, the generalized constructions based on the ``Faure 1992" and the ``Offset" factors randomized with a digital shift, however, the value of $n$ does not impact the results.

The results from this numerical experiment can also be somewhat explained by the criterion values in Tables \ref{tab:critF1} and \ref{tab:critH3}. If the $c$ values are high, that suggests a ``bad" point set that cannot be fixed via a shift, as a digital shift has little effect on the quality of a point set. However, even a ``bad" point set can be fixed with scrambling in a base $b$ such that $C_b \leq 1$.
\hchristiane{We need to explain better this paragraph. We should provide specific examples.}
Since the values of $n$ in Tables \ref{tab:critF1} and \ref{tab:critF3} are much smaller than the values of $n$ used in the convergence plots, not all the behaviour can be explained by the criterion values.
The ``Regular" Halton sequence, which had the highest $c$ values, had the worst performance by far when randomized with a digital shift. The ``Faure 1992" and ``Offset" Halton sequences, which had lower $c$ values, had RQMC errors almost as good as the scrambled Halton sequence, even when randomized with a digital shift.

\subsection{Comparing results using histograms}
\label{sec:hist}

To compare the use of deterministic permutations versus random scrambling, we use histograms to compare the integration error of the functions $h_0$ and $h_1$ obtained using specific generalized Halton and Faure sequences--- randomized via a digital shift unless otherwise specified---with the error distribution obtained using a base $\BFb$ random linear scrambling for the Halton sequence and a base $b$ nested scrambling for the Faure sequence. The point of this experiment is as follows: if a specific construction can easily be ``beaten'' by (have a larger error than) a randomized sequence obtained through scrambling, then this would be an argument in favor of using scrambling to improve the sequence instead of relying on a specific generalized sequence construction. If a construction consistently ranks better than most scrambled sequences, then this would be an argument to use those instead of scrambled sequences.

We consider three values of $s$: 4, 12, and 52. For the Faure sequence, the constructing base is the smallest prime larger or equal to $s$. Here, it is always equal to $s + 1$.

The reported error is the Mean Squared Error (MSE), and is estimated using $V = 25$ replications and 
a sample size of $n = 10000$, except in the $s = 52$ case where $n \in \{2809, 10000\}$.
We chose to visualize the $n = 2809, s = 52$ case as in Section \ref{sec:convergence}, the scrambled Faure sequence did not seem to outperform the generalized Faure sequence.


The results visualized via the histograms can be thought of taking a "slice" of the results from Section \ref{sec:convergence} at specific values of $n$. 
Note that the value of $n = 10000$ is not explicitly used in the convergence plots. 

For the Faure sequence, we compare the distribution of randomly scrambled point sets (as shown by the histogram) with the same three types of deterministically chosen factors as in Section \ref{sec:convergence} (original, F92 and F Offset), as well as the Monte Carlo methods. 

Similarly for the Halton sequence, we compare the distribution of randomly scrambled point sets (as shown by the histogram) with the same three types of deterministic permutations as in Section \ref{sec:convergence}, and also with the factors proposed in \cite{donglemieux22} (which we label ``DL'') and the ones propsoed in \cite{faure2009generalized} (which we label ``FL'').
\hchristiane{We need to explain what is DL and FL and then say we're also including those and not just the three from before}

Some of the values (namely ``Regular Shifted'' and Monte Carlo) obtained are sometimes quite large compared to the rest of the results. Therefore to better visualize the results, these large values have been excluded from the plots and their error values are reported in the caption instead. 

\statement{
\begin{figure}[htb]
\centering
\begin{subfigure}{0.5\textwidth}
\includegraphics[width=\linewidth]{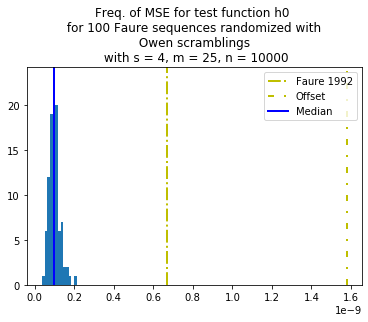}
\caption{Faure: 9.516e-10, MC: 3.879e-06}
\end{subfigure}%
\begin{subfigure}{0.5\textwidth}
\includegraphics[width=\linewidth]{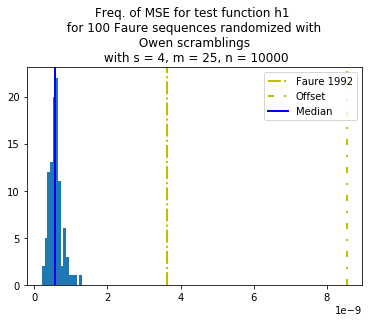}
\caption{Faure: 5.144e-09, MC: 2.207e-05}
\end{subfigure}
 \medskip
 \begin{subfigure}{0.5\textwidth}
 \includegraphics[width=\linewidth]{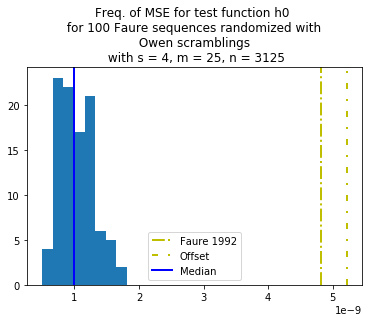}
 \caption{Faure: 4.2863477253700036e-09 \\
 MC: 1.2408852355892932e-05}
 \end{subfigure}%
 \begin{subfigure}{0.5\textwidth}
 \includegraphics[width=\linewidth]{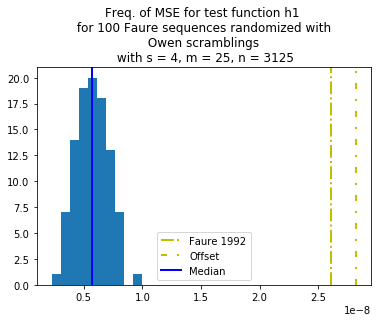}
 \caption{Faure: 2.322660364143871e-08 \\
 MC: 7.06337903694249e-05}
 \end{subfigure}
\caption{Comparing Nested Scrambling of the Faure sequence with other Faure sequences with $s = 4$}
\label{fig:FaureHist_s4}
\end{figure}

\begin{figure}[htb]
\centering
\begin{subfigure}{0.5\textwidth}
\includegraphics[width=\linewidth]{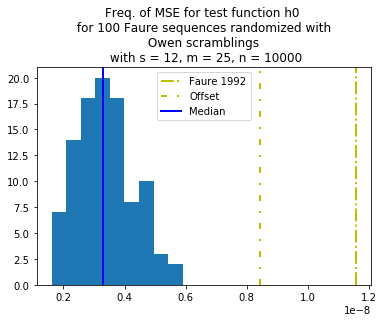}
\caption{Faure: 3.912e-08, MC: 1.163e-05}
\end{subfigure}%
\begin{subfigure}{0.5\textwidth}
\includegraphics[width=\linewidth]{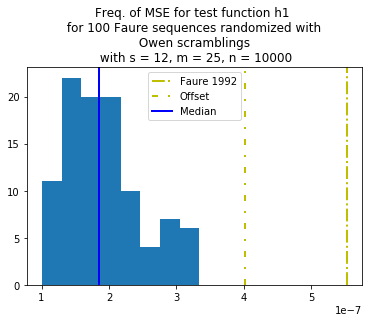}
\caption{Faure: 1.86e-06, MC: 0.000584}
\end{subfigure}
 \medskip
 \begin{subfigure}{0.5\textwidth}
 \includegraphics[width=\linewidth]{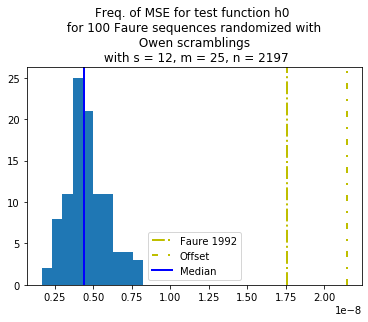}
 \caption{Faure: 1.913972257703249e-08 \\
 MC: 5.329731615648977e-05}
 \end{subfigure}%
 \begin{subfigure}{0.5\textwidth}
 \includegraphics[width=\linewidth]{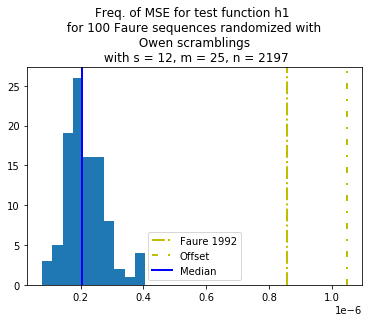}
 \caption{Faure: 9.321766608455894e-07 \\
 MC: 0.002683581619636522}
 \end{subfigure}
\caption{Comparing Nested Scrambling of the Faure sequence with other Faure sequences with $s = 12$}
\label{fig:FaureHist_s12}
\end{figure}

\begin{figure}[htb]
\centering
\begin{subfigure}{0.5\textwidth}
\includegraphics[width=\linewidth]{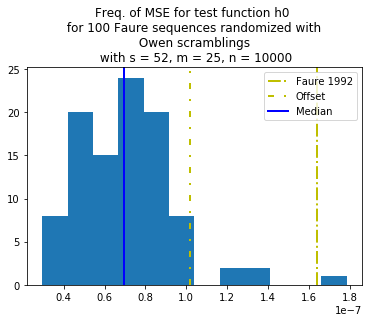}
\caption{Faure: 8.763e-07, MC: 5.0247e-05}
\end{subfigure}%
\begin{subfigure}{0.5\textwidth}
\includegraphics[width=\linewidth]{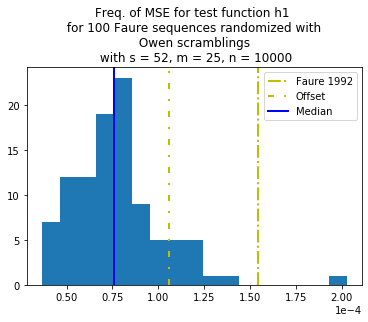}
\caption{Faure: 0.0006621, MC: 0.04699}
\end{subfigure}
\medskip
\begin{subfigure}{0.5\textwidth}
\includegraphics[width=\linewidth]{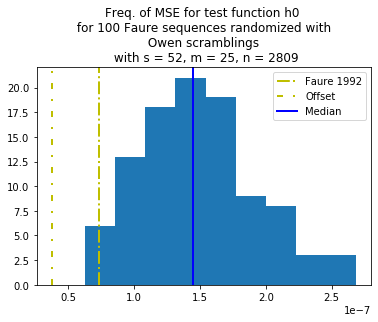}
\caption{Faure: 4.573e-08, MC: 0.0001795}
\end{subfigure}%
\begin{subfigure}{0.5\textwidth}
\includegraphics[width=\linewidth]{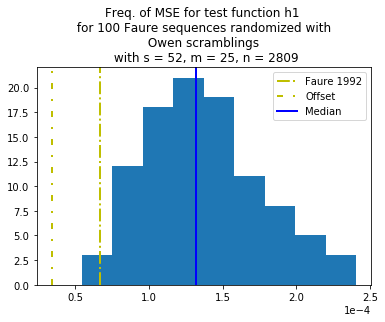}
\caption{Faure: 4.230e-05, MC: 0.1677}
\end{subfigure}
\caption{Comparing nested Scrambling of the Faure sequence with other Faure sequences with $s = 52$}
\label{fig:FaureHist_s52}
\end{figure}

\begin{figure}[htb]
\centering
\begin{subfigure}{0.5\textwidth}
\includegraphics[width=\linewidth]{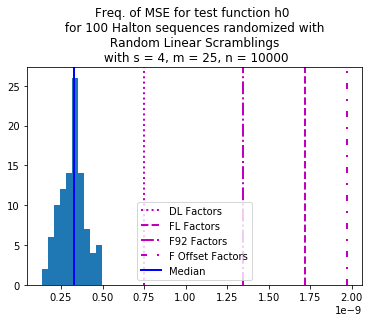}
\caption{Halton: 1.8421e-09, MC: 3.8787e-06}
\end{subfigure}%
\begin{subfigure}{0.5\textwidth}
\includegraphics[width=\linewidth]{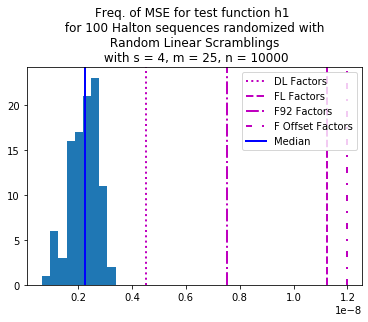}
\caption{Halton: 8.4177e-09, MC: 2.2075e-05}
\end{subfigure}
 \medskip
 \begin{subfigure}{0.5\textwidth}
 \includegraphics[width=\linewidth]{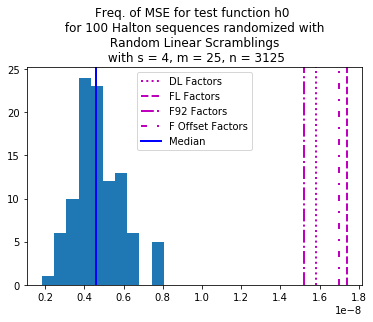}
 \caption{Halton: 1.81e-08 \\
 MC: 1.24e-05}
 \end{subfigure}%
 \begin{subfigure}{0.5\textwidth}
 \includegraphics[width=\linewidth]{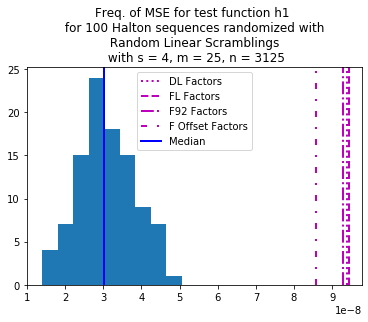}
 \caption{Halton: 1.09e-07 \\
 MC: 7.06e-05}
 \end{subfigure}
\caption{Comparing Random Linear Scrambling of the Halton sequence with other Halton sequences with $s = 4$}
\label{fig:HaltonHist_s4}
\end{figure}

\begin{figure}[htb]
\centering
\begin{subfigure}{0.5\textwidth}
\includegraphics[width=\linewidth]{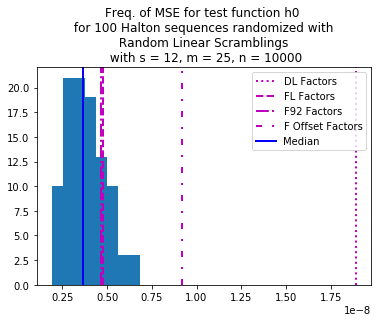}
\caption{Halton: 2.979e-08, MC: 1.1629e-05}
\end{subfigure}%
\begin{subfigure}{0.5\textwidth}
\includegraphics[width=\linewidth]{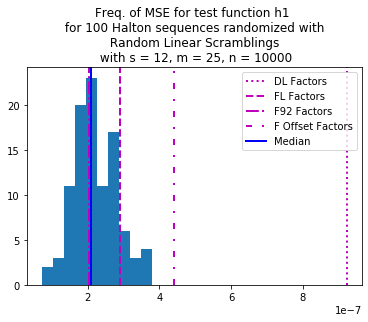}
\caption{Halton: 1.541e-06, MC: 0.000584}
\end{subfigure}
 \medskip
 \begin{subfigure}{0.5\textwidth}
 \includegraphics[width=\linewidth]{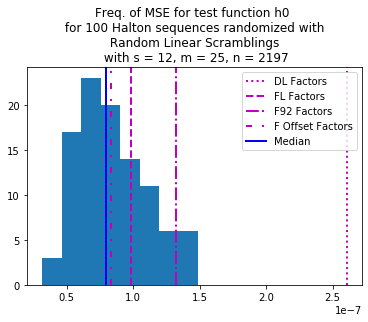}
 \caption{Halton: 1.05e-06, MC: 5.33e-05}
 \end{subfigure}%
 \begin{subfigure}{0.5\textwidth}
 \includegraphics[width=\linewidth]{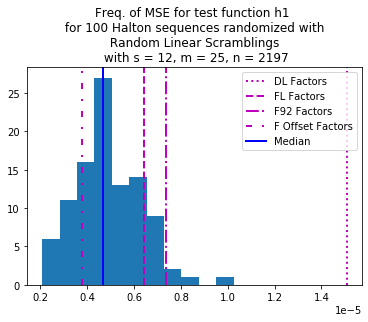}
 \caption{Halton: 4.52e-05, MC: 0.00268}
 \end{subfigure}
\caption{Comparing Random Linear Scrambling of the Halton sequence with other Halton sequences with $s = 12$}
\label{fig:HaltonHist_s12}
\end{figure}

\begin{figure}[htb]
\centering
\begin{subfigure}{0.5\textwidth}
\includegraphics[width=\linewidth]{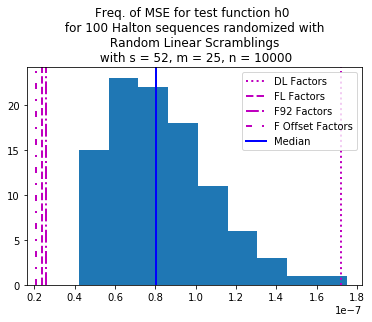}
\caption{Halton: 7.02e-06, MC: 5.024e-05}
\end{subfigure}%
\begin{subfigure}{0.5\textwidth}
\includegraphics[width=\linewidth]{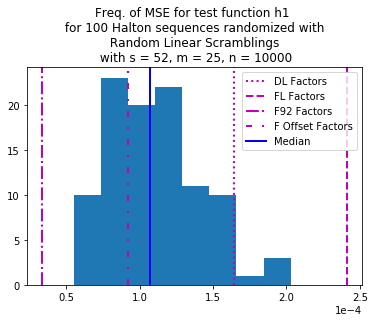}
\caption{Halton: 0.006982, MC: 0.04699}
\end{subfigure}
 \medskip
 \begin{subfigure}{0.5\textwidth}
 \includegraphics[width=\linewidth]{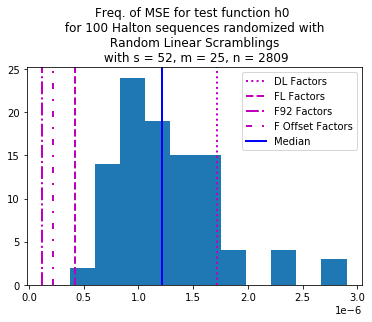}
 \caption{Halton: 4.89e-05, MC: 1.79e-04}
 \end{subfigure}%
 \begin{subfigure}{0.5\textwidth}
 \includegraphics[width=\linewidth]{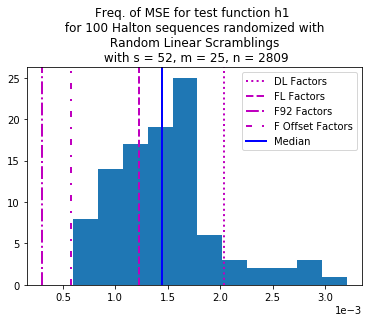}
 \caption{Halton: 0.0493, MC: 0.168}
 \end{subfigure}
\caption{Comparing Random Linear Scrambling of the Halton sequence with other Halton sequences with $s = 52$}
\label{fig:HaltonHist_s52}
\end{figure}
}

For the Faure sequence, the results suggest that in almost all cases, scrambling outperforms generalizing and then randomizing with a digital shift. The exception is the case where $s = 52, b = 53, n = 2809$, where generalizing and then applying a shift is more accurate (i.e., in the lower quartile of the histogram) than scrambling, but this result cannot be extrapolated into the general case. When $n$ is a small power of the constructing base $b$, the digital shift is more likely to perform well compared to scrambling.
For the Halton sequence, as well, in general, random linear scrambling is slightly better than using well-chosen permutations. However, since each of the dimensions are constructed in a different prime base, the choice of $n$ does not have a large impact on relative performance. 
This suggests that well chosen permutations have the potential to do better than scrambling but probably need to be chosen based on the type of function being integrated. Hence if one is looking for a multi-purpose construction, scrambling appears as a better choice.

\section{Conclusion}


In this paper, we evaluated two choices for multiplicative factors (or permutations) used to generalize the Halton and Faure sequences based on the algorithm from Faure \cite{faure1992good}.
We evaluated their performance using a numerical study as well as studying the $C_b(\BFk, P_n)$ values introduced in \cite{wiart2021dependence}. For Halton sequences, this algorithm shows a significant improvement on the corresponding results obtained in \cite{donglemieux22} for the cases where $s$ is larger than 4, as easily seen on the figures, especially with the ``Faure Offset" factors.
 \hchristiane{It's true that in most cases, the F Offset are better than DL, but when $s=4$ DL is the best. Also, random scrambling (as proposed in \cite{donglemieux22}) seems better than deterministic scrambling except in high dimensions, so I think this statement needs to be made more precise} 
These last factors are deterministic digital shifts  
and hence, at least when $n$ is a small power, they are a positive answer to the main question raised in the abstract and the introduction. 
However, random scrambling as proposed in \cite{donglemieux22} generally outperforms the use of deterministic scrambling, except in the $s=52$ case.
The constructions of Halton and Faure are radically different: Halton sequences grow step by step by adding the next prime from the list of increasing primes, contrary to Faure sequences that are built independently of the preceding ones, still with identity. Until now, the results for these sequences are not satisfactory compared to using random scrambling
\hchristiane{I think here as well we need to be more precise with this statement: I think we are trying to say that the F92 and F Offset factors do not work so well, and that random scrambling is a better choice in this case}
and a goal for future work is to assess different kinds of generating matrices in the framework of negative dependence based on \cite{faure1992good}.

\hchristiane{Element that could be added to the conclusion: the fact that the evaluation based on the $C_b(\BFk)$ values is consistent with the results obtained, i.e. for Halton, as $s$ increases, the GHAlton sequences have relatively smaller $c(2)$ than Halton, but for GFaure, their $c(2)$ is not smaller than Faure}

\section*{Acknowledgements}

The third author gratefully acknowledges the financial support of NSERC via grant \#238959.


\begin{thebibliography}{10}

\bibitem{asotsky2003one}
D.I. Asotsky and I.M. Sobol'.
\newblock One more experiment on estimating high-dimensional integrals by
  quasi-monte carlo methods.
\newblock {\em Mathematics and Computers in Simulation}, 62(3-6):255--263,
  2003.

\bibitem{donglemieux22}
G.Y. Dong and C.~Lemieux.
\newblock Dependence properties of scrambled {H}alton sequences.
\newblock {\em Mathematics and Computers in Simulation}, 2022.
\newblock To appear.

\bibitem{Faure81}
H.~Faure.
\newblock Discr\'epance des suites associ\'ees \`a un syst\`eme de num\'eration
  (en dimension un).
\newblock {\em Bull. Soc. Math. France}, 109:143--182, 1981.

\bibitem{faure1982}
H.~Faure.
\newblock Discr{\'e}pance de suites associ{\'e}es {\`a} un syst{\`e}me de
  num{\'e}ration (en dimension s).
\newblock {\em Acta Arithmetica}, 41(4):337--351, 1982.

\bibitem{faure1992good}
H.~Faure.
\newblock Good permutations for extreme discrepancy.
\newblock {\em Journal of Number Theory}, 42(1):47--56, 1992.

\bibitem{faure2009generalized}
H.~Faure and C.~Lemieux.
\newblock Generalized {H}alton sequences in 2008: A comparative study.
\newblock {\em ACM Transactions on Modeling and Computer Simulation (TOMACS)},
  19(4):1--31, 2009.

\bibitem{FKP}
H.~Faure,  P.~Kritzer, and F.~Pillichshammer.
\newblock From van der {C}orput to modern constructions of sequences for
  quasi-{M}onte {C}arlo rules.
\newblock {\em Indag.\ Math.}, 26:760--822, 2015.

\bibitem{rHAL60a}
J.~H. Halton.
\newblock On the efficiency of certain quasi-random sequences of points in
  evaluating multi-dimensional integrals.
\newblock {\em Numerische Mathematik}, 2:84--90, 1960.

\bibitem{joyboyletan},
C.~Joy, P.~Boyle and K.-S.~Tan
\newblock Quasi-Monte Carlo methods in numerical
finance.
\newblock {\em Manag. Sci.}, 42(6):926--938, 1996.

\bibitem{lemieux2009mcqmc}
C.~Lemieux.
\newblock {\em Monte Carlo and Quasi-Monte Carlo Sampling}.
\newblock Springer Series in Statistics. Springer New York, 2009.

\bibitem{Lem17}
C.~Lemieux.
\newblock Negative dependence, scrambled nets, and variance bounds.
\newblock {\em Mathematics of Operations Research}, 43(1):228--251, 2018.

\bibitem{qFAU09a}
C.~Lemieux and H.~Faure.
\newblock New perspectives on $(0,s)$-sequences.
\newblock In P.~L'Ecuyer and A.B. Owen, editors, {\em Monte Carlo and
  Quasi-Monte Carlo Methods 2008}, pages 113--130. Springer-Verlag, 2009.

\bibitem{unanchored2022}
C.~Lemieux and J.~Wiart.
\newblock On the distribution of scrambled $(0,m,s)-$nets over unanchored
  boxes.
\newblock In A.~Keller, editor, {\em Monte Carlo and Quasi-Monte Carlo Methods
  2020}. Springer, 2022.

\bibitem{matousek1998thel2}
J.~Matous\v{e}k.
\newblock On the {L}$_2$-discrepancy for anchored boxes.
\newblock {\em Journal of Complexity}, 14(4):527--556, 1998.

\bibitem{okten2009generalized}
G.~{\"O}kten.
\newblock Generalized von {N}eumann--{K}akutani transformation and random-start
  scrambled halton sequences.
\newblock {\em Journal of Complexity}, 25(4):318--331, 2009.

\bibitem{owen1995randomly}
A.B. Owen.
\newblock Randomly permuted $(t, m, s)$-nets and $(t, s)$-sequences.
\newblock In {\em {M}onte {C}arlo and quasi-{M}onte {C}arlo methods in
  scientific computing}, pages 299--317. Springer, 1995.
  
\bibitem{owensobol}
A.B. Owen.
\newblock Scrambling Sobol' and Niederreiter-Xing points.
\newblock {\em Journal of Complexity}, 14(4):466--489, 1998.

\bibitem{owen2017randomized}
A.B. Owen.
\newblock A randomized {H}alton algorithm in R.
\newblock {\em arXiv preprint arXiv:1706.02808}, 2017.

\bibitem{owen2020dropping}
A.B. Owen.
\newblock On dropping the first Sobol' point.
\newblock In: Keller, A. (eds) {\em Monte Carlo and Quasi-Monte Carlo Methods. MCQMC 2020}. Springer Proceedings in Mathematics \& Statistics, 387:71--86, 2022.

\bibitem{BeatMC}
A. Papageorgiou and J. Traub.
\newblock Beating Monte Carlo.
\newblock {\em Simulation Risk, Vol 9, No 6, June 1996}.



\bibitem{sobol1967distribution}
I.M. Sobol'.
\newblock On the distribution of points in a cube and the approximate
  evaluation of integrals.
\newblock {\em Zhurnal Vychislitel'noi Matematiki i Matematicheskoi Fiziki},
  7(4):784--802, 1967.

\bibitem{rTEZ95a}
S.~Tezuka.
\newblock {\em Uniform Random Numbers: Theory and Practice}.
\newblock Kluwer Academic Publishers, Norwell, MA, 1995.

\bibitem{rTEZ94b}
S.~Tezuka and T.~Tokuyama.
\newblock A note on polynomial arithmetic analogue of {H}alton sequences.
\newblock {\em ACM Transactions on Modeling and Computer Simulation},
  4:279--284, 1994.

\bibitem{wiart2021dependence}
J.~Wiart, C.~Lemieux, and G.Y. Dong.
\newblock On the dependence structure and quality of scrambled $(t, m,
  s)$-nets.
\newblock {\em Monte Carlo Methods and Applications}, 2021.

\end{thebibliography}

\end{document}